\documentclass{webofc}
\usepackage[varg]{txfonts}   
\begin{document}
\title{On the widths and binding energies of $K^-$ nuclear states \\ 
and the role of $K^-$ multi-nucleon interactions}
%
%

\author{\firstname{Jaroslava} \lastname{Hrt\'ankov\'a}\inst{1}\fnsep\thanks{\email{hrtankova@ujf.cas.cz}} \and
        \firstname{Ji\v{r}\'{\i}} \lastname{Mare\v{s}}\inst{1}\fnsep\thanks{\email{mares@ujf.cas.cz}} 
}

\institute{Nuclear Physics Institute, 250 68 \v{R}e\v{z}, Czech Republic  
          }

\abstract{%
We report on our recent self-consistent calculations of $K^-$ nuclear quasi-bound states using 
$K^-$ optical potentials derived from chirally motivated meson-baryon coupled channels 
models~\cite{hmPLB17,hmPRC17}. The 
$K^-$ single-nucleon potentials were supplemented by a phenomenological $K^-$ multi-nucleon 
interaction term introduced to achieve good fits to $K^-$ atom  data. We demonstrate a substantial impact 
of the $K^-$ multi-nucleon absorption on the widths of $K^-$ nuclear states. If such states ever exist in 
nuclear many-body systems, their widths are excessively large to allow observation.

}
\maketitle
\section{Introduction}

The near-threshold $K^-N$ attraction generated by the $\Lambda(1405)$ resonance seems to be strong 
enough to allow binding of the $K^-$ meson in nuclei. This fact stimulated numerous theoretical studies of 
$K^-$ nuclear quasi-bound states as well as experimental searches for $K^-$ nuclear clusters~\cite{W10,ghm16,nvs17}. 
However, the question of their binding energies and widths, and even their very existence is still 
far from being settled. The $K^-N$ interaction near threshold has been recently described within 
chirally motivated coupled channels meson-baryon interaction models~\cite{kmnlo,pnlo,m,b}, parameters 
of which have been 
fitted to low-energy experimental data~\cite{mNPB,sid}. $K^-N$ scattering amplitudes which 
enter the construction of 
$K^-$ optical potentials are strongly energy dependent near and below threshold, i.e., in the region  
relevant for calculations of kaonic nuclear states. It is important to treat this energy 
dependence properly~\cite{cfggmPLB,cfggmPRC11}. 
The above chiral meson-baryon interaction models include only $K^-$  absorption on a single-nucleon, 
$K^-N \rightarrow \pi Y$ ($Y=\Sigma, \Lambda$).   
However, in the nuclear medium, $K^-$ multi-nucleon interactions take place as well~\cite{mfgPLB,fgmNPA,gmNPA}, and 
have to be considered in realistic evaluations of $K^-$ widths (and to lesser extent also $K^-$ binding 
energies) in the nuclear medium. Recently, Friedman and Gal~\cite{fgNPA16} supplemented $K^-$ single-nucleon 
potentials constructed within several chiral coupled channels meson-baryon models by a phenomenological 
term representing the $K^-$ multi-nucleon interactions. They fitted parameters of this phenomenological term 
to kaonic data for each meson-baryon interaction model separately. Only the P~\cite{pnlo} and KM~\cite{kmnlo} models 
were able to reproduce both kaonic data and experimental fractions of $K^-$ absorption at 
rest~\cite{bubble1,bubble2,bubble3}.      
In this contribution we apply these two models to calculations of $K^-$ nuclear quasi-bound states in selected 
nuclei.   

In Section 2, we briefly introduce kaon self-energy operator including in-medium modifications of the underlying 
$K^-N$ scattering amplitudes. We discuss how to incorporate strong energy dependence of the in-medium 
amplitudes in self-consistent calculations of kaonic nuclear quasi-bound states. In Section 3, we present 
selected results of our calculations aiming at demonstrating the crucial role played by $K^-$ multi-nucleon 
interactions in the nuclear medium. A brief summary in Section 4 concludes this contribution.  

\section{Methodology}

The interaction of the $K^-$ meson with a nucleus is described by the Klein-Gordon equation

\begin{equation}\label{KG}
 \left[ \vec{\nabla}^2  + \tilde{\omega}_{K^-}^2 -m_{K^-}^2 -\Pi_{K^-}(\omega_{K^-},\rho) \right]\phi_{K^-} = 0~,   
\end{equation}
where, $\tilde{\omega}_{K^-} = m_{K^-} - B_{K^-} -{\rm i}\Gamma_{K^-}/2 -V_C= \omega_{K^-} - V_C$ with 
$B_{K^-}$ ($\Gamma_{K^-}$) being the $K^-$ binding energy (width), $m_{K^-}$ denotes the $K^-$ mass, $V_C$ is 
the Coulomb potential, and  $\rho$ is the nuclear density distribution. The kaon self-energy  
$\Pi_{K^-}$ is constructed from in-medium isospin 0 and 1 $s$-wave amplitudes $F_0$ and $F_1$ in a $t\rho$ form as follows: 
\begin{equation}\label{piK}
\Pi_{K^-} = 2\text{Re}( {\omega}_{K^-})V_{K^-}^{(1)}=-4\pi \frac{\sqrt{s}}{m_N}\left(F_0\frac{1}{2}\rho_p + F_1\left(\frac{1}{2}\rho_p+\rho_n\right)\right)~,
\end{equation}
where  $V_{K^-}^{(1)}$ is the $K^-$-nucleus optical potential, $m_N$ is the nucleon mass, and the kinematical factor 
$\sqrt{s}/m_N$ transforms the scattering amplitudes from the two-body frame to the $K^-$-nucleus frame. 
The proton and neutron density distributions, $\rho_p$ and $\rho_n$, are evaluated within the relativistic mean-field model NL-SH~\cite{nlsh}. 

The in-medium scattering amplitudes were obtained from the free-space amplitudes $f_{K^- p}$ and $f_{K^- n}$, derived within 
chiral coupled channels meson-baryon interaction models P and KM, using the multiple scattering approach 
(WRW)~\cite{wrw} in order to account for Pauli correlations. They are of the following form:
\begin{equation}\label{Eq.:in-med amp}
F_{1}=\frac{f_{K^-n}(\sqrt{s})}{1+\frac{1}{4}\xi_k \frac{\sqrt{s}}{m_N} f_{K^-n}(\sqrt{s}) \rho}~, \quad F_{0}=\frac{[2f_{K^-p}(\sqrt{s})-f_{K^-n}(\sqrt{s})]}{1+\frac{1}{4}\xi_k \frac{\sqrt{s}}{m_N}[2f_{K^-p}(\sqrt{s}) - f_{K^-n}(\sqrt{s})] \rho}~,
\end{equation}
where $\xi_k$ is adopted from Ref.~\cite{fgNPA16}.

\begin{figure}[t!]
\centering
\includegraphics[width=0.48\textwidth,clip]{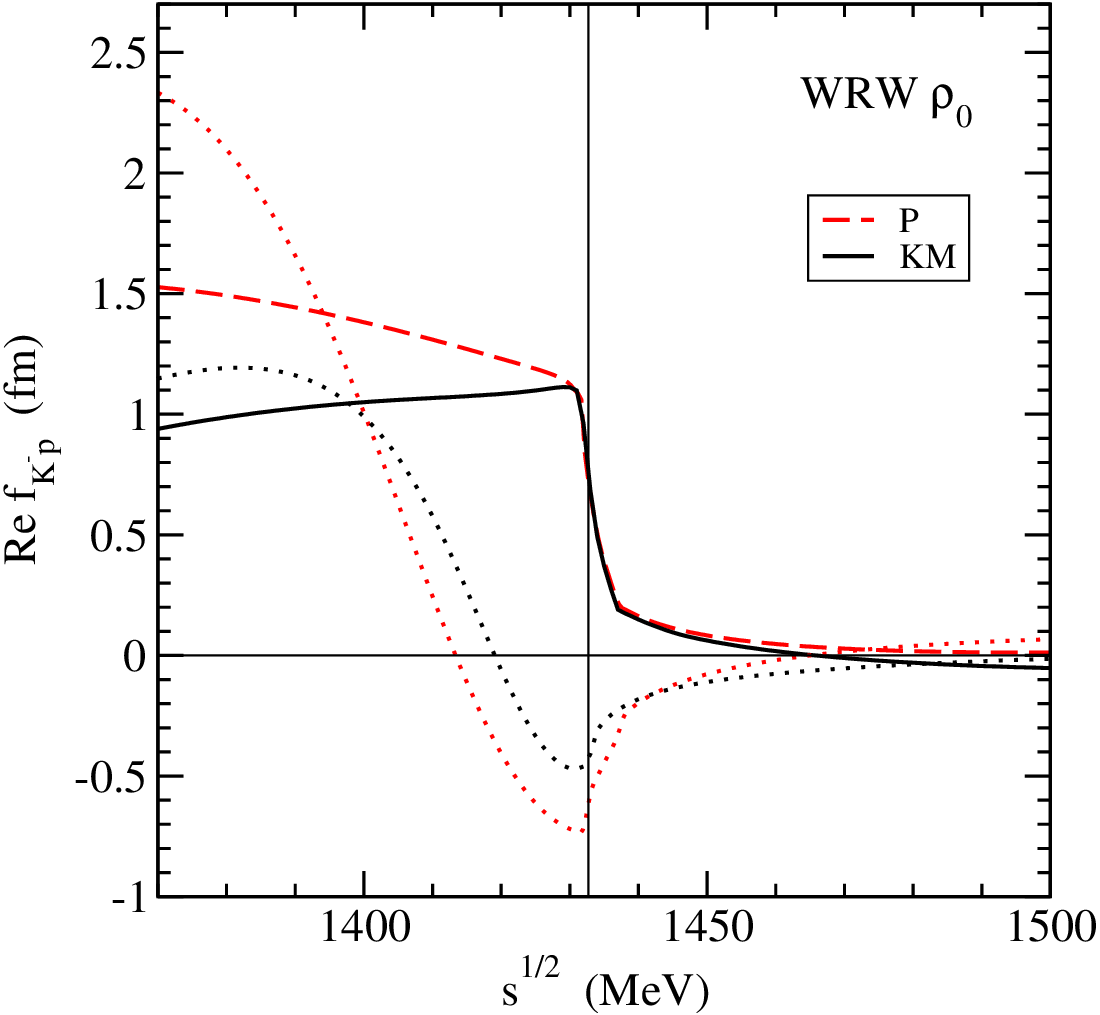}\hspace*{5pt}
\includegraphics[width=0.48\textwidth,clip]{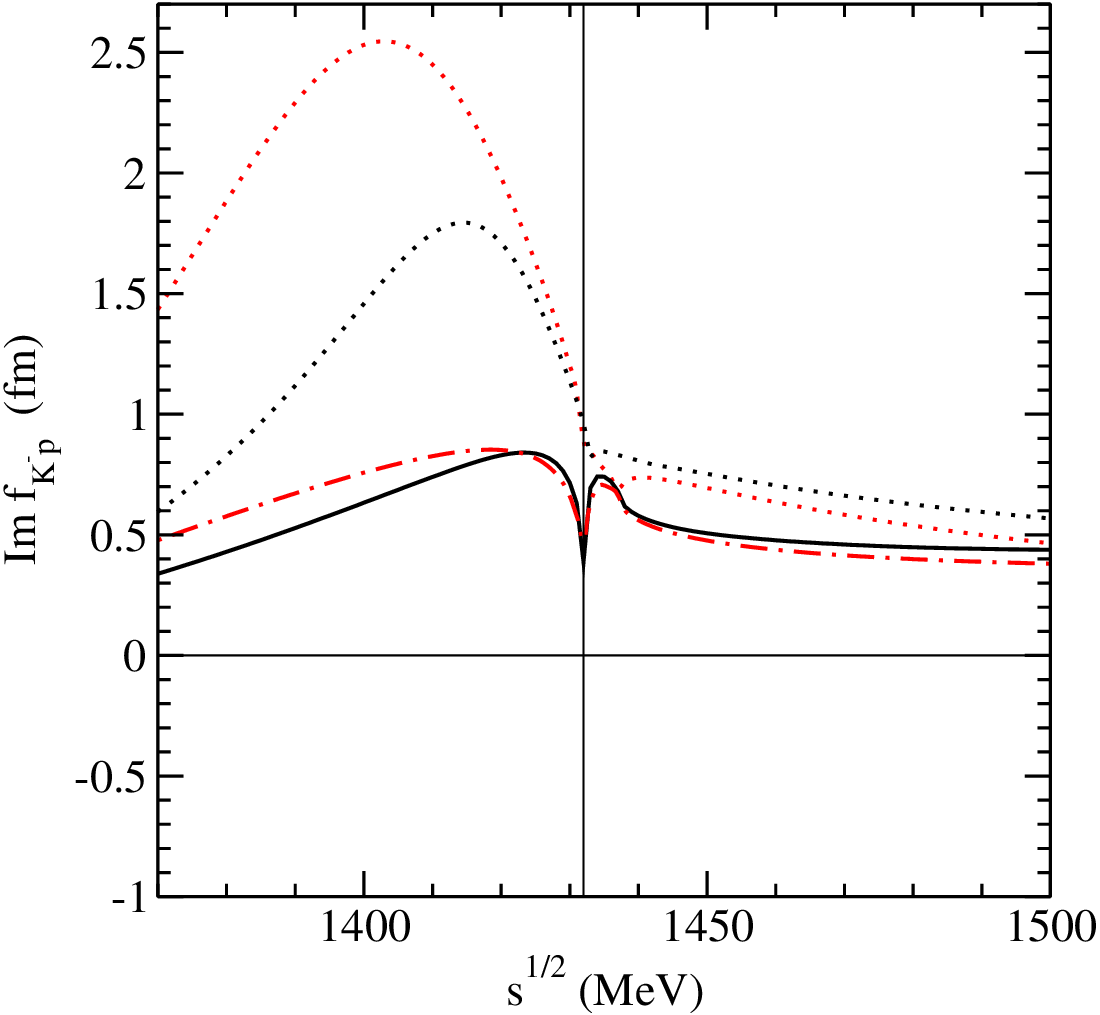}\\[10pt]
\includegraphics[width=0.48\textwidth,clip]{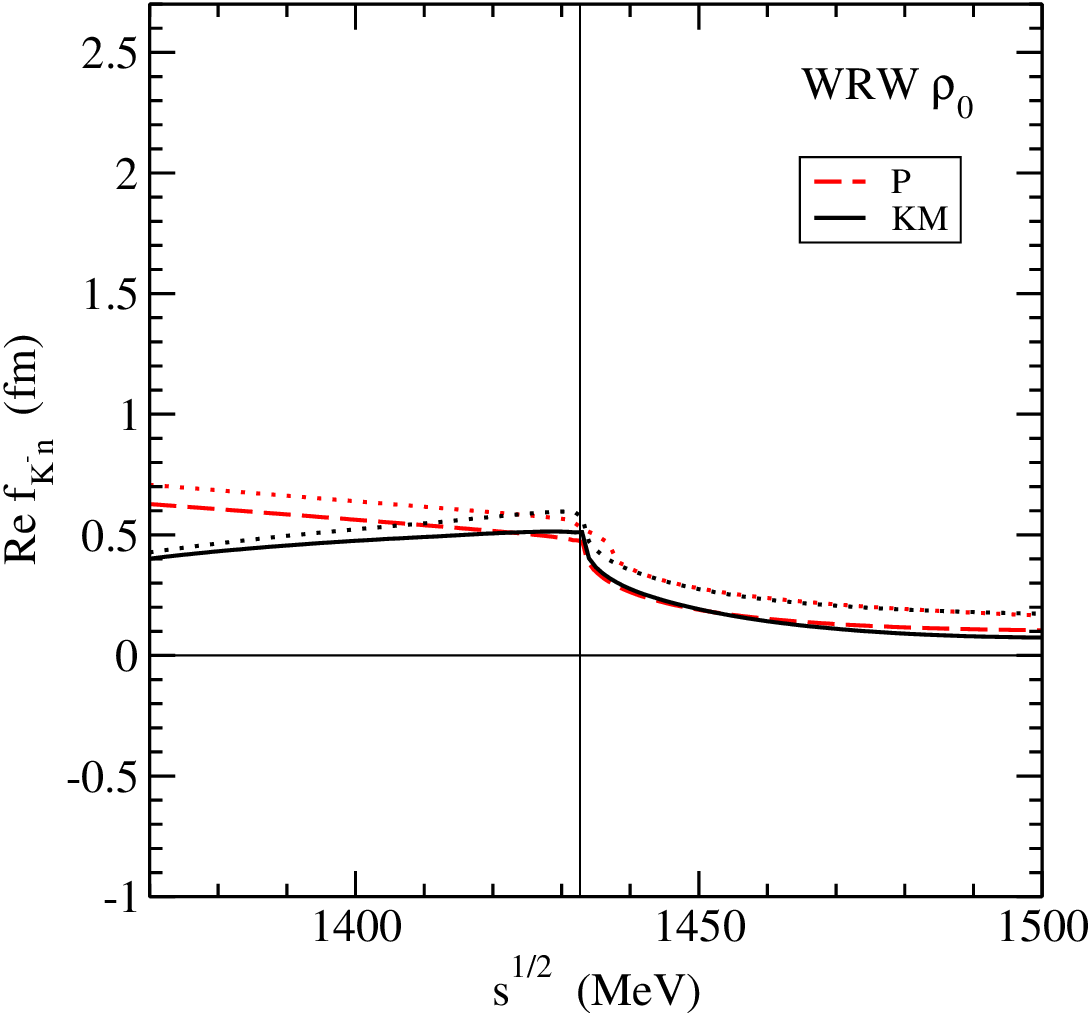}\hspace*{5pt}
\includegraphics[width=0.48\textwidth,clip]{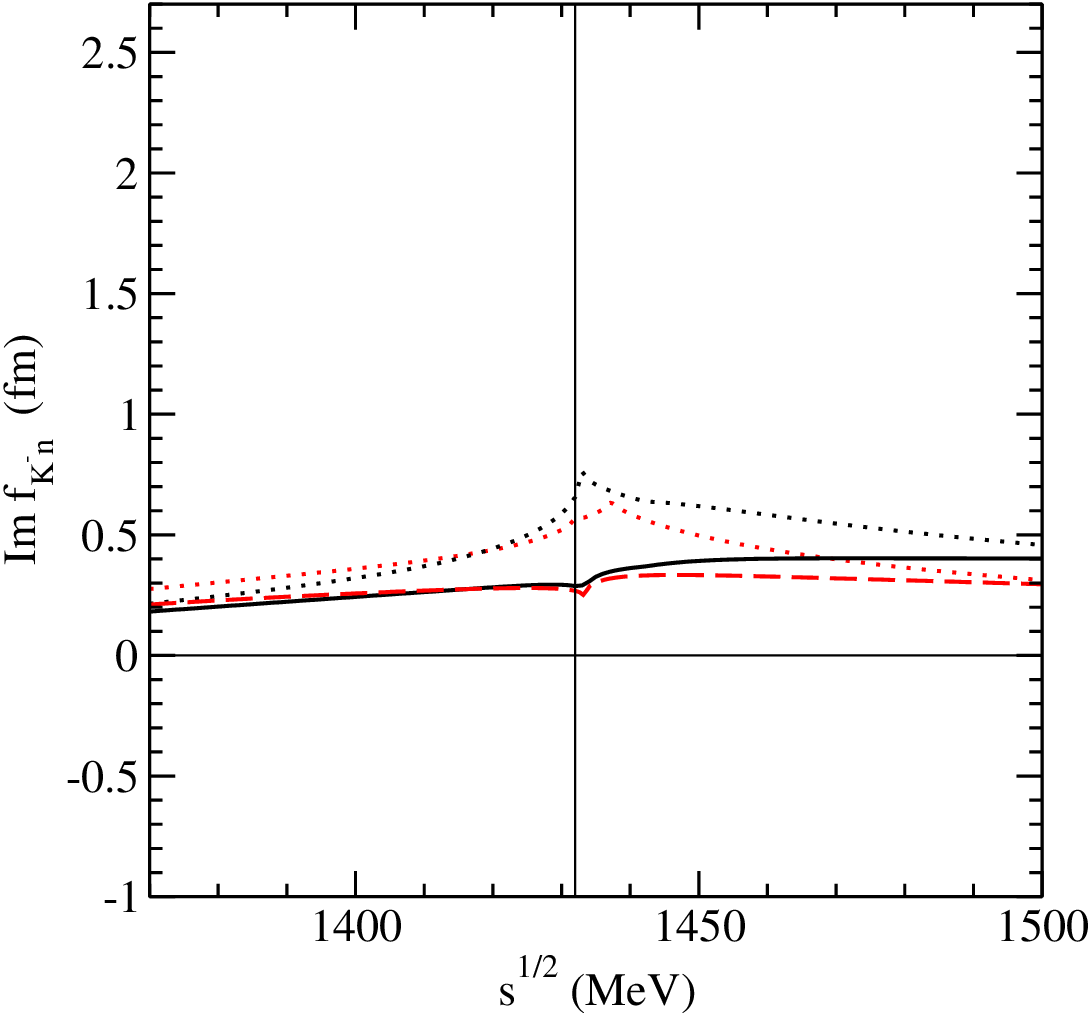}
\caption{
Energy dependence of real (left) and imaginary (right) parts of WRW modified $K^-p$ (top) and $K^- n$ (bottom) amplitudes at $\rho_0=0.17$~fm$^{-3}$ in the P and KM models. 
Corresponding free-space amplitudes are shown for comparison (dotted lines).  
}
\label{fig-1}       
\end{figure}
\begin{figure}[t!]
\centering
\includegraphics[width=0.48\textwidth,clip]{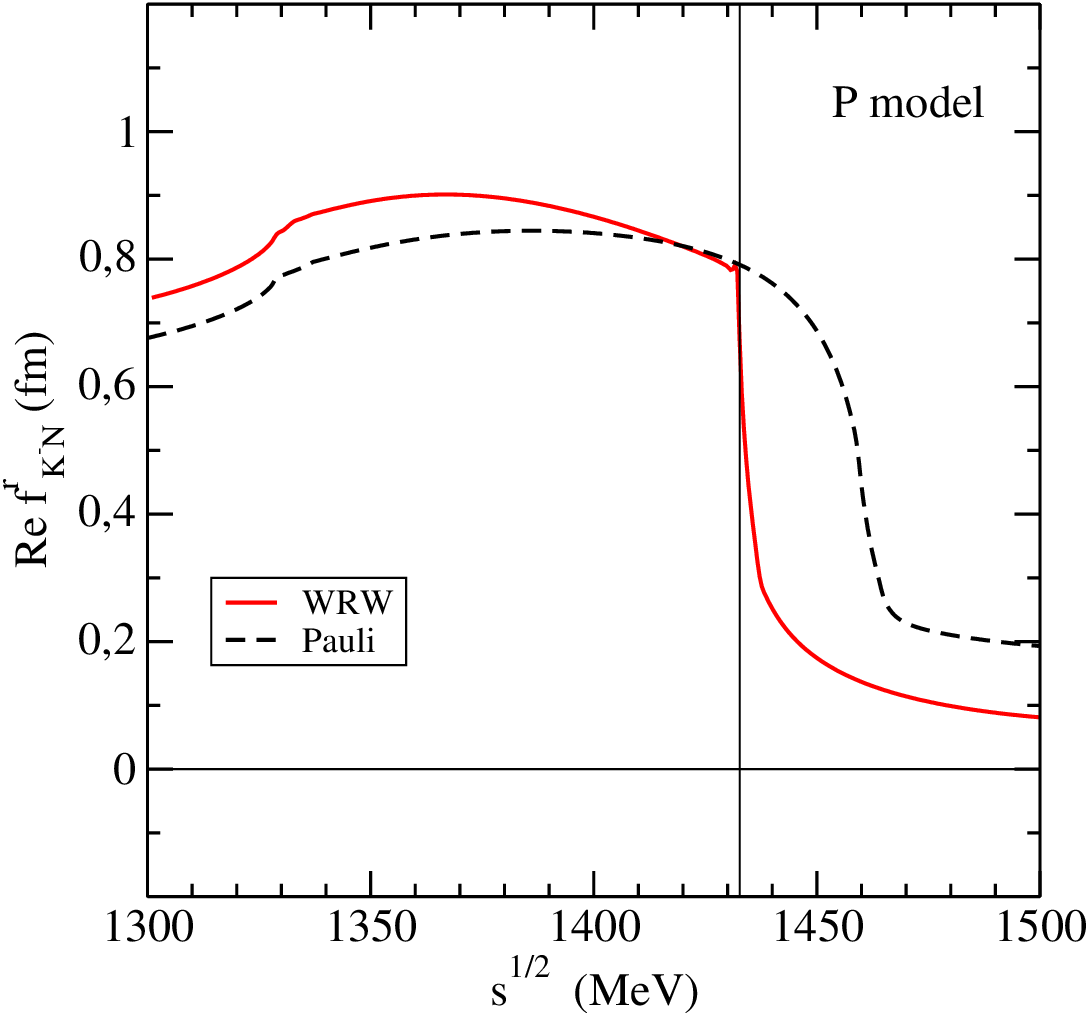}\hspace*{5pt}
\includegraphics[width=0.48\textwidth,clip]{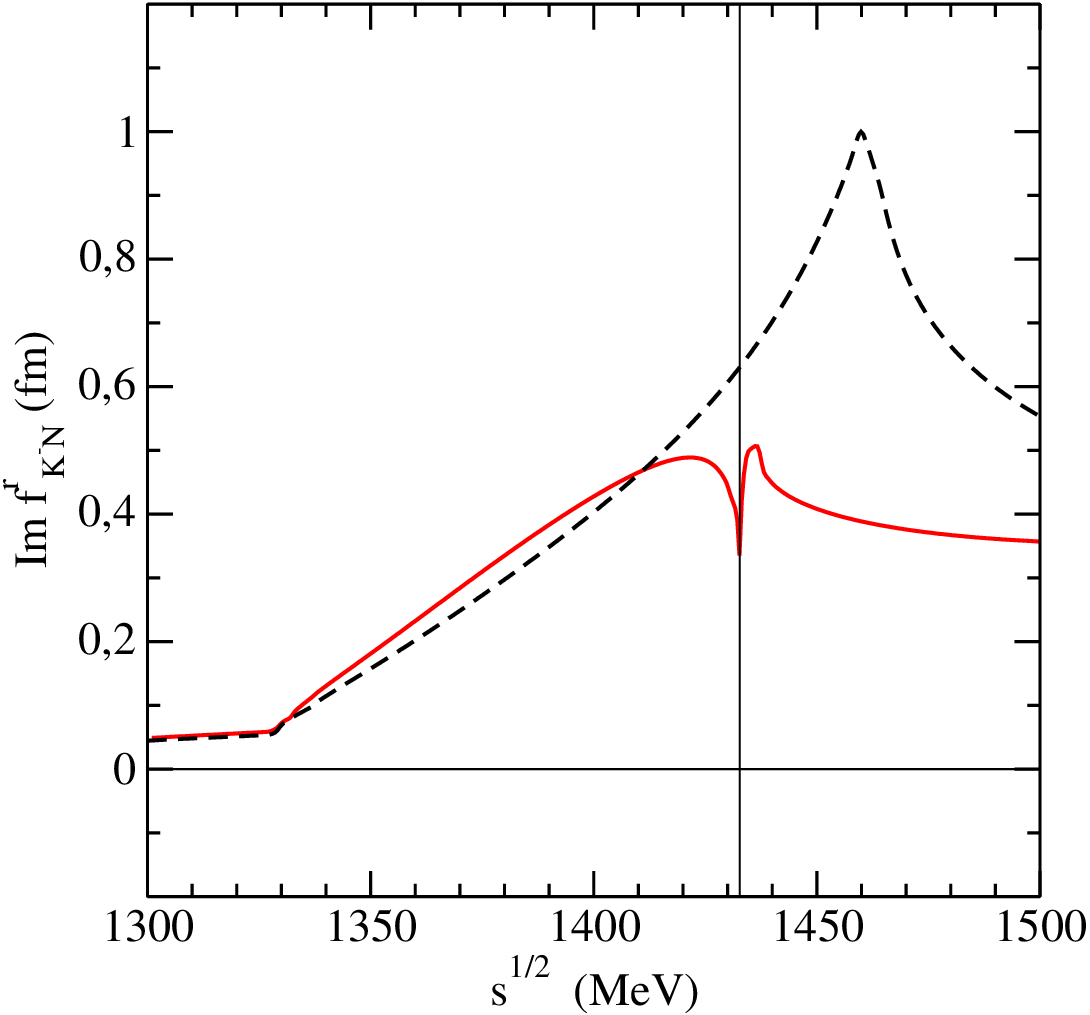}
\caption{
Comparison of real (left) and imaginary (right) parts of WRW (full line) and Pauli (dashed line) modified 
reduced amplitudes 
$f^{\rm r}_{K^- N}=\frac{1}{2}(f^{\rm r}_{K^- p}+f^{\rm r}_{K^- n})$ plotted as function of energy $\sqrt{s}$ 
for $\rho_0=0.17$~fm$^{-3}$ in the P model (see text for details). 
}
\label{fig-2}       
\end{figure}

In Fig.~\ref{fig-1}, we present the $K^-p$ (top) and $K^-n$  (bottom) amplitudes in the P and KM models, 
modified by the WRW procedure at saturation density $\rho_0=0.17$~fm$^{-3}$, plotted as a function of energy. 
The corresponding free scattering amplitudes (dotted lines) are presented for comparison.   
The $K^-p$ in-medium amplitudes are affected significantly by Pauli correlations: the real parts of the 
amplitudes become attractive in the entire energy region below threshold, and the imaginary parts are considerably 
lowered. On the other hand, the $K^-n$ amplitudes are modified by Pauli correlations only moderately. 

In previous calculations~\cite{cfggmPLB, gmNPA}, the in-medium $K^- N$ amplitudes in the P model~\cite{pnlo} were constructed 
in a different way. The integration over the intermediate meson-baryon momenta in the underlying Green's function was restricted to a region ensuring the nucleon intermediate energy to be above the Fermi level (denoted further 'Pauli'). 
In Fig.~\ref{fig-2}, we compare the Pauli correlated amplitudes with the WRW 
modified amplitudes in the P model. It is rewarding that both approaches, WRW and Pauli, yield similar $K^-N$ in-medium reduced 
amplitudes\footnote{$f^{\rm r}_{K^-N} = g(p)f_{K^-N}g(p')$, where $g(p)$ is a momentum-space form factor 
(see Ref.~\cite{cfggmPLB})} $f^{\rm r}_{{K^-}N}=\frac{1}{2}(f^{\rm r}_{K^- p}+f^{\rm r}_{K^- n})$ in the subthreshold energy region which is 
relevant to our calculations of $K^-$ nuclear states. Above threshold, the behavior of Pauli and WRW modified 
amplitudes is different. 

Important feature of the $K^- p$ scattering amplitudes shown in Figs. 1 and 2 is their strong energy dependence near 
threshold caused by the presence of the $\Lambda(1405)$ resonance, dynamically generated in chirally motivated 
coupled channels models.    
This feature calls for a proper self-consistent treatment during the evaluation of the $K^-$ optical potential. 
The amplitudes are a function of energy defined by the Mandelstam variable  
\begin{equation}
 s=(E_N+E_{K^-})^2-(\vec{p}_N+\vec{p}_{K^-})^2~,
\end{equation}
where  $E_N=m_N-B_N$, $E_{K^-}=m_{K^-}-B_{K^-}$ and $\vec{p}_{N(K^-)}$ is the nucleon (kaon) momentum. 
In the $K^-$-nucleus frame the momentum dependent term is no longer zero and provides additional downward energy shift. 
The energy shift $\delta \sqrt{s} = \sqrt{s} - E_{th} = \sqrt{s} - (m_N +m_K)$ was expanded in terms of binding and kinetic energies as 
follows \cite{hmPLB17,hmPRC17}: 
\begin{equation} \label{Eq.:deltaEs}
 \delta \sqrt{s}=  -B_N\frac{\rho}{\bar{\rho}}\, - \beta_N\! \left[B_{K^-}\frac{\rho}{\rho_{\rm max}} + T_N\left(\frac{\rho}{\bar{\rho}}\right)^{2/3}\!\!\!\! +V_C\left(\frac{\rho}{\rho_{\rm max}}\right)^{1/3}\right] + \beta_{K^-} {\rm Re}V_{K^-}(r)~,
\end{equation}
where $B_N=8.5$~MeV is the average binding energy per nucleon, $\bar{\rho}$ is the average nuclear density, $\rho_{\rm max}$ is the maximal value of the nuclear density, $\beta_{N(K^-)}={m_{N(K^-)}}/(m_N+m_{K^-})$ and $T_N=23$~MeV is the average nucleon kinetic energy in the Fermi Gas Model. 
The energy shift respects the low-density limit, i.~e. $\delta \sqrt{s} \rightarrow 0$ as $\rho \rightarrow 0$. 

The above P and KM models describe only the $K^-$ interactions with a single nucleon. However, $K^-$ interactions with 
two and more nucleons take place in the medium as well~\cite{mfgPLB,fgNPA16}. Therefore, we supplemented the 
$K^-$ single-nucleon potential $V_{K^-}^{(1)}$ from Eq.~\eqref{piK} with a phenomenological optical potential 
$V_{K^-}^{(2)}$ describing the $K^-$ multi-nucleon interactions: 
\begin{equation} \label{Vknn}
 2\text{Re}(\omega_{K^-})V_{K^-}^{(2)}=-4 \pi B (\frac{\rho}{\rho_0})^{\alpha} \rho~.
\end{equation}

The parameters of the phenomenological potential, complex amplitude $B$ and positive exponent $\alpha$ listed in 
Table~\ref{tab-1}, were recently fitted by Friedman and Gal~\cite{fgNPA16} to kaonic atom data for each $K^-N$ 
interaction model separately. 
Moreover, the total $K^-$ optical potential, $V_{K^-}=V_{K^-}^{(1)}+V_{K^-}^{(2)}$, was confronted 
with branching ratios of $K^-$ absorption at rest from bubble chamber experiments \cite{bubble1, bubble2, bubble3} and 
the P and KM models discussed in this contribution were found to be the only two interaction models capable of 
reproducing both experimental constraints simultaneously. 
\begin{table}[b!]
\centering
\caption{
Values of the complex amplitude $B$ and exponent $\alpha$ used to evaluate $V_{K^-}^{(2)}$  for chiral
meson-baryon interaction models considered in this work.
}
\label{tab-1}       
 \begin{tabular}{|c|cc|cc|}
 \hline
  & & & & \\[-10pt]
  & P1  & KM1 & P2 & KM2  \\ \hline \hline
$\alpha$   & 1 & 1 & 2 & 2\\
Re$B$ (fm) &  $-$1.3 $\pm$ 0.2 & $-$0.9 $\pm$ 0.2 & $-$0.5 $\pm$ 0.6 & 0.3 $\pm$ 0.7 \\
Im$B$ (fm) &  ~~1.5 $\pm$ 0.2 & ~~1.4 $\pm$ 0.2 & ~~
4.6 $\pm$ 0.7 & 3.8 $\pm$ 0.7\\ 
\hline
 \end{tabular}
\vspace*{10pt}  
\end{table}

The kaonic atom data probe the $K^-$ optical potential up to at most $\sim 50$\% of the nuclear density~\cite{fgNPA16}.  
Further inside the nucleus, the shape of the potential is just a matter of extrapolation to higher densities. 
Therefore, we considered two options in our calculations: a) we applied the formula~\eqref{Vknn} in the entire nucleus 
(full density option - FD) 
b) we fixed the potential $V_{K^-}^{(2)}$ at constant value $V_{K^-}^{(2)}(0.5\rho_0)$  for $\rho (r) \ge 0.5 \rho_0$ 
(half density limit - HD).

\section{Results}
The formalism outlined in the previous section was adopted to self-consistent calculations of $K^-$ nuclear 
quasi-bound states in various nuclei across the periodic table. Here we present only few selected results, for 
more details see Refs.~\cite{hmPLB17,hmPRC17}).  

\begin{figure}[h!]
\centering
\includegraphics[width=0.52\textwidth,clip]{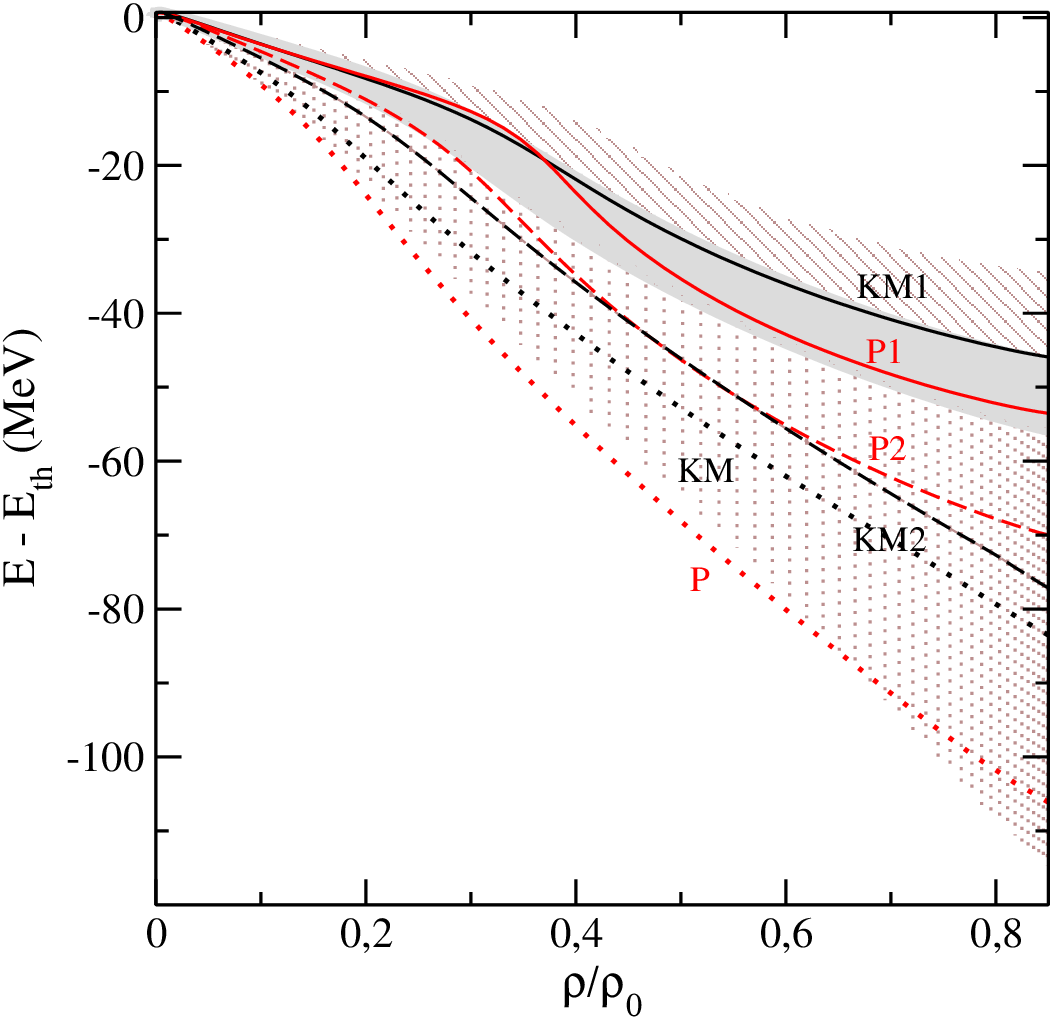}
\caption{
Subthreshold energies probed in the $^{208}$Pb$+K^-$ nucleus
as a function of relative density $\rho/ \rho_0$, calculated self-consistently in the single-nucleon $K^-N$ amplitude models P and KM (dotted lines), 
supplemented by the FD variant of $V_{K^-}^{(2)}$ with $\alpha=1$ (full lines) and $\alpha=2$ (dashed lines). The shaded areas stand for uncertainties.
}
\label{fig-3}       
\end{figure}

The self-consistently evaluated energy shift $\delta \sqrt{s}$ from Eq~(\ref{Eq.:deltaEs}) is strongly density 
dependent which plays important role in calculations of kaonic nuclear, as well as atomic states.    
Figure~\ref{fig-3} illustrates the strong density dependence of $\delta \sqrt{s}$ 
in $^{208}$Pb, calculated self-consistently within the P and KM models 
augmented by the FD variant of $V^{(2)}_{K^-}$ with $\alpha=1$ (P1, KM1) 
and $\alpha=2$ (P2, KM2).  The P1 and KM1 models yield smaller energy shift with respect to the 
$K^- N$ threshold than the original single-nucleon potentials $V_{K^-}^{(1)}$ (P and KM). 
The P2 and KM2 models yield energy shifts closer to the single-nucleon potentials.  
The energy shift for the FD option is in any case shallower than for 
the original $K^-$ single-nucleon potential owing to very strong absorption. 
It is to be noted that the P and KM models supplemented with various versions of the multi-nucleon interaction 
term could be regarded as equivalent since they all lie in corresponding uncertainty bands and describe 
kaonic atom data equally well.

\begin{figure}[t!]
\centering
\includegraphics[width=0.49\textwidth,clip]{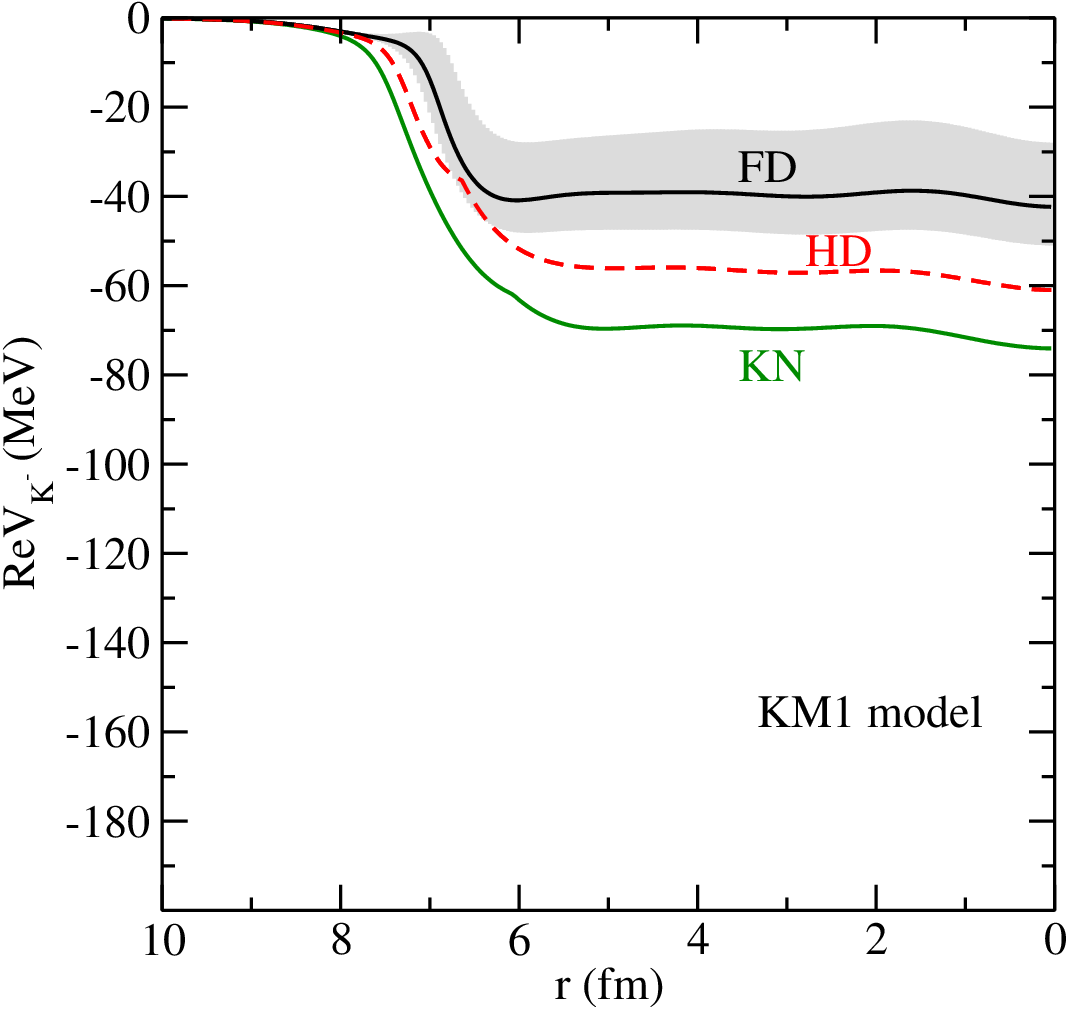}\hspace*{-7pt}
\includegraphics[width=0.49\textwidth,clip]{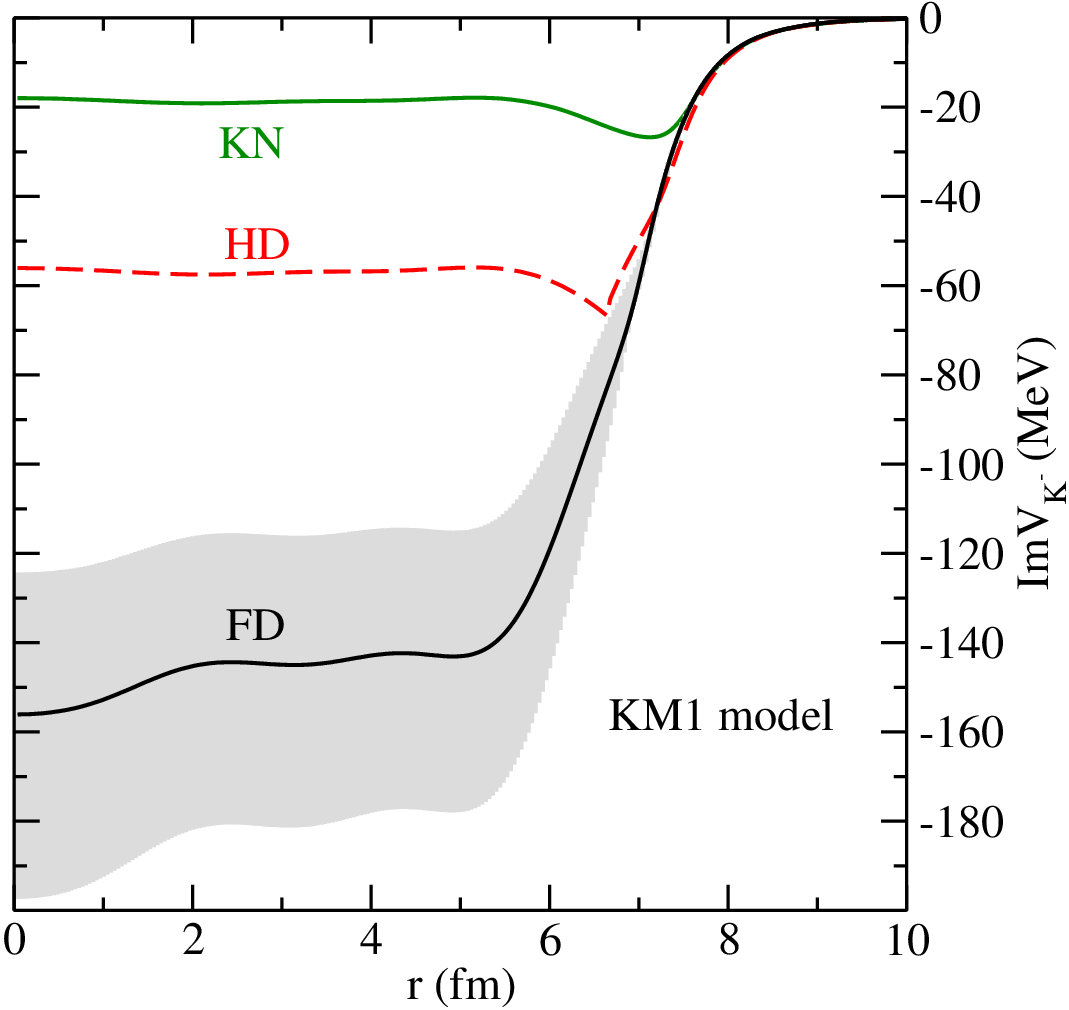}\\[10pt]
\includegraphics[width=0.49\textwidth,clip]{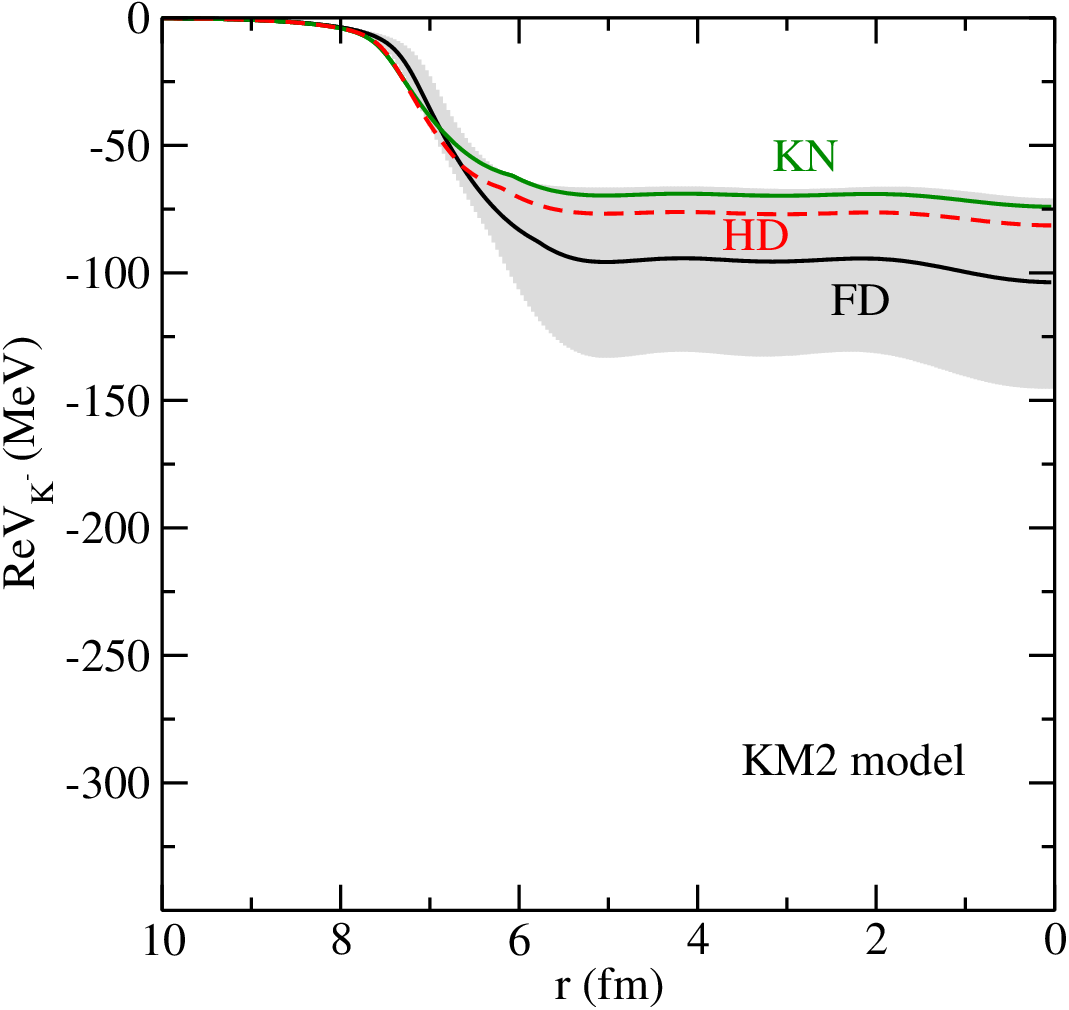}\hspace*{-7pt}
\includegraphics[width=0.49\textwidth,clip]{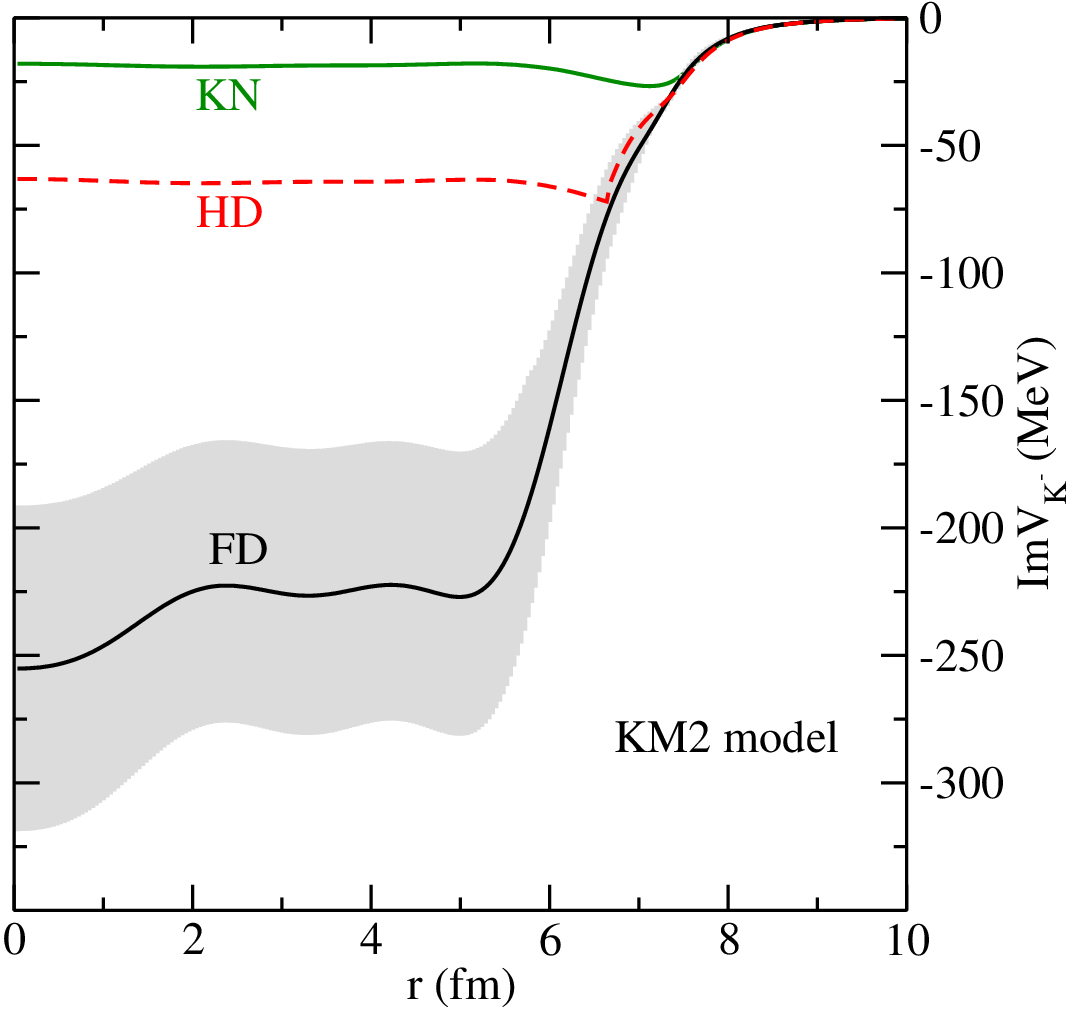}
\caption{The real (left) and imaginary (right) part of the $K^-$ optical potential in 
$^{208}$Pb, calculated self-consistently in the KM1 (top) and KM2 (bottom) model,
for the HD and FD versions of the $K^-$ multi-nucleon potential (see text for details). The shaded areas stand 
for uncertainties. The single-nucleon $K^-$ potential (KN, green solid lines) calculated in the KM model is shown for comparison.}
\label{fig-4} 
\end{figure}

In Fig.~\ref{fig-4}, we present real (left) and imaginary (right) parts of the total $K^-$ potential in  $^{208}$Pb, 
calculated self-consistently in the KM1 and KM2 models for the HD and FD options of the multi-nucleon 
potential $V_{K^-}^{(2)}$. 
The gray shaded areas stand for uncertainties in $V_{K^-}^{(2)}$ due to 
the complex parameter $B$ (see Table~\ref{tab-1}). 
The underlying chirally-inspired $K^-$ single-nucleon potential (denoted by `KN') are shown for comparison.
The real parts of the $K^-$ optical potential are affected by multi-nucleon interactions markedly less than its  
imaginary parts in all considered models. In the KM1 model, the FD and HD options yield Re$V_{K^-}$ shallower than 
the original single-nucleon $V_{K^-}^{(1)}$ potential. The same holds for the P1 and P2 models (not shown in the figure). 
In the KM2 model, the overall $K^-$ real potential is deeper than the underlying $K^-$ single-nucleon potential due 
to the positive sign of Re$B$. 
The Re$V_{K^-}$ potentials for HD and FD options differ between each other up to $\approx 20$~MeV in each interaction model.  
On the other hand, the imaginary parts of $V_{K^-}$ exhibit much larger dispersion for different versions 
of $V_{K^-}^{(2)}$, as illustrated in Fig.~\ref{fig-4}, right panels. The $K^-$ multi-nucleon absorption 
significantly deepens the imaginary part of the $K^-$ optical potential. For the FD option of $V_{K^-}^{(2)}$, 
the KM model yields $\mid$Im$V_{K^-}\!\!\mid \gg \mid$Re$V_{K^-}\!\!\mid$ inside the nucleus for both values of $\alpha$, even when the uncertainties of the $K^-$ multi-nucleon potential are taken into account. The same holds for the P model (not shown in the figure).   

\begin{figure}[t!]
\centering
\includegraphics[width=0.52\textwidth,clip]{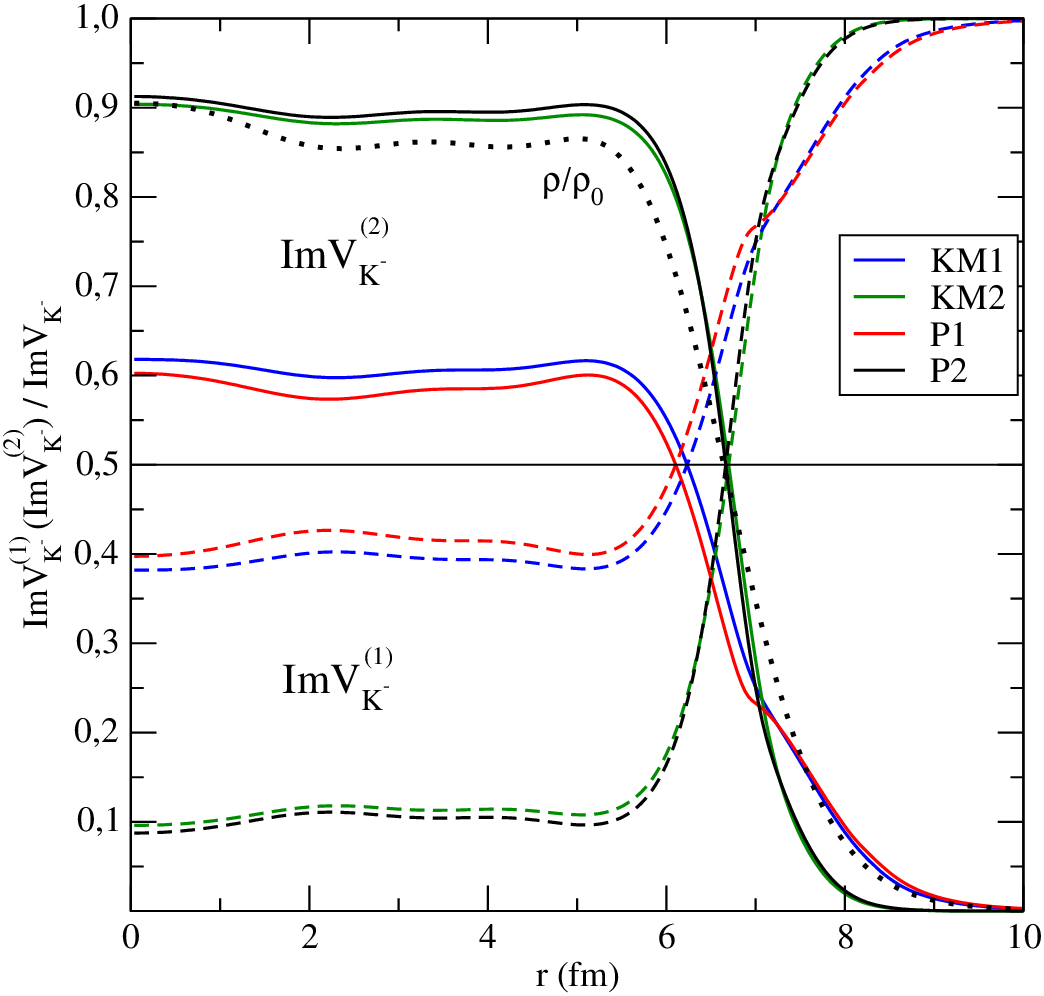}
\caption{
The ratio of Im$V_{K^-}^{(1)}$ (dashed line) and Im$V_{K^-}^{(2)}$ (solid line) potentials to the
total $K^-$ imaginary potential Im$V_{K^-}$ as a function of radius, calculated self-consistently
for $^{208}$Pb$+K^-$ system  within different $K^-N$ interaction models and FD option for the
$K^-$ multi-nucleon potential. The relative nuclear density $\rho / \rho_0$ (dotted line)  is shown for comparison.
}
\label{fig-5}       
\end{figure}

The particular role of $K^-$ single- and multi-nucleon absorptions with respect to the nuclear density is 
illustrated in Fig.~\ref{fig-5}. Here we compare individual contributions of $K^-$ single nucleon and 
multi-nucleon absorptions to the total $K^-$ absorption, expressed as a fraction of Im$V_{K^-}^{(1)}$ and 
Im$V_{K^-}^{(2)}$ with respect to the total imaginary $K^-$ potential Im$V_{K^-}$ in  $^{208}$Pb, 
calculated self-consistently within the P and KM models for the FD option of $V_{K^-}^{(2)}$. 
The density $\rho / \rho_0$ (thin dotted line) is shown for comparison. 
 Since the range of the corresponding potentials is different, the relative contributions of 
Im$V_{K^-}^{(1)}$ and Im$V_{K^-}^{(2)}$ to the $K^-$ absorption are changing with radius (density). 
At the nuclear surface, the $K^-$ absorption on a single nucleon dominates, while it is reduced in the nuclear 
interior due to vicinity of $\pi \Sigma$ threshold and the multi-nucleon absorption prevails.
The models with $\alpha=2$ yield lower relative fraction of the single-nucleon $K^-N$ absorption in the nuclear medium 
than the models with $\alpha=1$. It is due to the self-consistent value of $\sqrt{s}$ at $\rho_0$ which is 
closer to the $K^-N \rightarrow \pi \Sigma$ threshold in the case with $\alpha=2$ (see Fig.~\ref{fig-3}).
It is to be noted that both options for multi-nucleon potential, HD and FD, yield similar fraction of the 
$K^-$ single-nucleon and multi-nucleon absorption inside the nucleus. 

\begin{table}[t!]
\centering
\caption{
1s $K^-$ binding energies and widths (in MeV) in selected nuclei calculated using the single-nucleon
$K^-N$ KM amplitudes (denoted KN); plus a phenomenological amplitude $B(\rho/\rho_0)^{\alpha}$, where $\alpha=1$ and 2,
for half-density limit (HD) and full density option (FD) (see text
for details).
}
\label{tab-2}       
 \begin{tabular}{|r c|c|c c|c c|}
\hline 
       \multicolumn{3}{|c}{}& \multicolumn{2}{|c|}{}& \multicolumn{2}{|c|}{} \\[-10pt]
   \multicolumn{3}{|c}{KM model}& \multicolumn{2}{|c|}{$\alpha = 1 $}& \multicolumn{2}{|c|}{$\alpha = 2$} \\ \hline
   &  &\hspace*{7pt} KN \hspace*{7pt}& \hspace*{7pt} HD  \hspace*{7pt} & \hspace*{7pt}FD \hspace*{7pt} & \hspace*{7pt} HD\hspace*{7pt} & \hspace*{7pt} FD \hspace*{7pt} \\ \hline 
      & & & & & & \\[-10pt]
   $^{12}$C  & $B_{K^-}$ & 45 & 34 & not & 48 & not  \\
      & $\Gamma_{K^-}$ & 44 & 114 & bound & 125  & bound \\ \hline
      & & & & & & \\[-10pt]
      $^{16}$O  & $B_{K^-}$ & 45 & 34 & not & 48 & not  \\
      & $\Gamma_{K^-}$ & 40 & 109 & bound & 121  & bound \\ \hline
      & & & & & & \\[-10pt]
      $^{40}$Ca  & $B_{K^-}$ & 59 & 50 & not & 64 & not  \\
      & $\Gamma_{K^-}$ & 37 & 113 & bound & 126  & bound \\ \hline
      & & & & & & \\[-10pt]
      $^{90}$Zr  & $B_{K^-}$ & 69 & 56 & 17 & 72 & 30 \\
      & $\Gamma_{K^-}$ & 36 & 107 & 312 & 120  & 499 \\ \hline
      & & & & & & \\[-10pt]
      $^{208}$Pb  & $B_{K^-}$ & 78 & $\; 64$ & 33 & 80 & 53  \\
      & $\Gamma_{K^-}$ & 38 & 108 & 273 & 122  & 429 \\ \hline 
          \multicolumn{5}{c}{}&\multicolumn{2}{c}{} \\[-5pt] 
      \hline  
       \multicolumn{3}{|c}{}& \multicolumn{2}{|c|}{}& \multicolumn{2}{|c|}{} \\[-10pt]
 \multicolumn{3}{|c}{P model}& \multicolumn{2}{|c|}{$\alpha = 1 $}& \multicolumn{2}{|c|}{$\alpha = 2$} \\ \hline
   &  & KN & HD & FD &  HD & FD \\ \hline 
      & & & & & & \\[-10pt]
   $^{12}$C  & $B_{K^-}$ & 64 & 50 & not & 64  & not  \\
      & $\Gamma_{K^-}$ & 28 & 96 & bound & 122  & bound \\ \hline
      & & & & & & \\[-10pt]
      $^{16}$O  & $B_{K^-}$ & 64 & 50 & not & 63 & not  \\
      & $\Gamma_{K^-}$ & 25 & 94 & bound & 117  & bound \\ \hline
      & & & & & & \\[-10pt]
      $^{40}$Ca  & $B_{K^-}$ & 81 & 67 & not & 82 & not  \\
      & $\Gamma_{K^-}$ & 14 & 95 & bound & 120  & bound \\ \hline
      & & & & & & \\[-10pt]
      $^{90}$Zr  & $B_{K^-}$ & 90 & 74 & 19 & 87 & not  \\
      & $\Gamma_{K^-}$ & 12 & 88 & 340 & 114  & bound \\ \hline
      & & & & & & \\[-10pt]
      $^{208}$Pb  & $B_{K^-}$ & 99 & 82 & 37 & 96 & $47^{*}$  \\
      & $\Gamma_{K^-}$ & 14 & 92 & 302 & 117  & $412^{*}$ \\
\hline
\end{tabular}
\vspace*{5pt}

* the solution of Eq.~\ref{KG} for Im$V_{K^-}$ scaled by factor 0.8 
\end{table}

Finally, in Table~\ref{tab-2} we present $1s$ $K^-$ binding energies $B_{K^-}$ and widths $\Gamma_{K^-}$, calculated 
within the P and KM models. 
For comparison, we show also $K^-$ binding energies and widths calculated only with the chirally-inspired 
$K^-$ single-nucleon potentials. 
The  $K^-$ multi-nucleon interactions cause radical increase of $K^-$ widths, while $K^-$ binding energies 
change much less. The  HD option of the $V_{K^-}^{(2)}$ potential yields $K^-$ widths of 
order $\sim 100$~MeV and the binding energies much smaller than the corresponding widths in most nuclei. 
The FD multi-nucleon potential $V_{K^-}^{(2)}$ even does not predict any antikaon bound state in the majority of nuclei. 
We found $1s$ $K^-$ quasi-bound states in $^{90}$Zr and $^{208}$Pb but the corresponding $K^-$ binding energies are small and 
the widths are one order of magnitude larger than the binding energies~\footnote{For the FD variant of the P2 model, we had to scale huge imaginary part Im$V_{K^-}$ by factor 0.8 in order to get fully converged self-consistent solution of 
Eq.~\ref{KG}.}. 

\section{Summary}
In this contribution, we reported on our most recent self-consistent calculations  of $K^-$ nuclear quasi-bound states 
performed using a $K^-$ single-nucleon potential derived within two chirally motivated 
coupled channels  meson-baryon interaction models P and KM, supplemented by a phenomenological potential representing  
$K^-$ multi-nucleon interactions. The applied models were recently fitted to kaonic atom data and confronted with 
the branching ratios of $K^-$ single-nucleon absorption at rest~\cite{fgNPA16}. 
We demonstrated that the $K^-$ multi-nucleon absorption gives rise to substantial increase of $K^-$ absorption widths. 
In vast majority of nuclei the widths of $K^-$ nuclear quasi-bound states exceed considerably corresponding 
$K^-$ binding energies. Identification of such states in experiment seems thus highly unlikely. 
 
\section*{Acknowledgements}
We thank E. Friedman and A. Gal for valuable discussions. 
This work was supported by the GACR Grant No. P203/15/04301S.

\end{document}